\documentclass[a4paper,10pt,twocolumn,fleqn]{article}

\usepackage[utf8]{inputenc}
\usepackage[T1]{fontenc}

\usepackage[british]{babel}

\usepackage[scale=0.88,columnsep=.8cm]{geometry}
\usepackage[font={small},labelfont={bf},labelsep=period]{caption}

\usepackage{amsmath,amssymb,mathtools}
\usepackage{newtxtext,newtxmath}
\usepackage{bm}
\usepackage{graphicx}
\usepackage{natbib}
\usepackage{aas_macros}
\usepackage{hyperref}

\newcommand{\degr}{\ensuremath{{}^\circ}}

\renewcommand{\vec}[1]{\bm{#1}}
\newcommand{\mat}[1]{\bm{\mathrm{#1}}}
\newcommand{\tp}{\mathrm{T}}
\newcommand{\D}{\mathrm{d}}
\newcommand{\E}{\mathrm{e}}
\newcommand{\I}{\mathrm{i}}
\DeclareMathOperator{\Ev}{E}
\DeclareMathOperator{\Var}{Var}
\DeclareMathOperator{\Cov}{Cov}
\DeclareMathOperator{\erfc}{erfc}
\newcommand{\SNR}{\mathrm{SNR}}
\newcommand{\ESNR}{\mathrm{\hat{S}NR}}
\renewcommand{\Re}{\operatorname{Re}}
\renewcommand{\Im}{\operatorname{Im}}

\begin{document}

\title{Moment-Based Ellipticity Measurement as a \\
Statistical Parameter Estimation Problem}

\author{%
Nicolas Tessore\thanks{}
\and Sarah Bridle
}

\date{{\small%
Jodrell Bank Centre for Astrophysics, University of Manchester, \\[-2pt]
Alan Turing Building, Oxford Road, Manchester, M13 9PL, UK}
\\[\baselineskip]
\today}

\twocolumn[
\begin{@twocolumnfalse}
\maketitle
\begin{abstract}
We show that galaxy ellipticity estimation for weak gravitational lensing with unweighted image moments reduces to the problem of measuring a combination of the means of three independent normal random variables.
Under very general assumptions, the intrinsic image moments of sources can be recovered from observations including effects such as the point-spread function and pixellation.
Gaussian pixel noise turns these into three jointly normal random variables, the means of which are algebraically related to the ellipticity.
We show that the random variables are approximately independent with known variances, and provide an algorithm for making them exactly independent.
Once the framework is developed, we derive general properties of the ellipticity estimation problem, such as the signal-to-noise ratio, a generic form of an ellipticity estimator, and Cram\'er-Rao lower bounds for an unbiased estimator.
We then derive the unbiased ellipticity estimator using unweighted image moments.
We find that this unbiased estimator has a poorly behaved distribution and does not converge in practical applications, but demonstrates how to derive and understand the behaviour of new moment-based ellipticity estimators.
\end{abstract}
\vskip 3\baselineskip
\end{@twocolumnfalse}
]

{
\renewcommand{\thefootnote}{\fnsymbol{footnote}}
\footnotetext[1]{%
Email: \url{nicolas.tessore@manchester.ac.uk}}
}

\section{Introduction}

Weak gravitational lensing is the method with the most potential to constrain the nature of dark energy \citep[e.g.\@][]{2006astro.ph..9591A,2006ewg3.rept.....P}.
However, it relies on extremely accurate measurement of the shapes of millions to billions of galaxies.
Numerous methods and challenges have been developed to try to meet this potential.
The first methods used quadrupole moments to estimate ellipticities \citep{1990ApJ...349L...1T,1995ApJ...449..460K,1995A&A...303..331B},
with later methods using basis functions or model fitting \citep{1999A&A...352..355K,2002sgdh.conf...38B,2007MNRAS.382..315M,2013MNRAS.434.1604Z}.
These shear estimates are then usually calibrated due to the biases found in shear measurement challenges \citep{2006MNRAS.368.1323H,2007MNRAS.376...13M,2009AnApS...3....6B,2010MNRAS.405.2044B,2010arXiv1009.0779K,2014ApJS..212....5M}.

Although model fitting can be made statistically rigorous, not propagating the resulting distributions through to cosmology can still lead to biased results \citep{2014MNRAS.441.2528K}.
Furthermore, if the models used are not representative of real galaxies, then an additional ``model bias'' must be corrected \citep{2010MNRAS.404..458V,2012MNRAS.424.2757M}.
Finally, model fitting is usually very slow, and takes up a significant fraction of the computing time required to obtain cosmology from current weak lensing surveys. 
Quadrupole moments, on the other hand, tend to be very noisy, unless a weighting function is used, which then might require a complicated scheme to try and correct the biases due to weighting \citep{1995ApJ...449..460K,2000ApJ...537..555K}.
Even so, these weighted quadrupole moment techniques have to be calibrated in practice, as they are not unbiased in a strict statistical sense.
A range of different shear calibration methods have been developed, and currently the most promising seems to be the metacalibration technique of resimulation using the SHERA algorithm \citep{2017arXiv170202600H,2012MNRAS.420.1518M}.

The fundamental principle of shear measurement is to observe the ellipticity of distant sources, which has been slightly altered by light deflection due to the large-scale structure of the universe.
For a source having elliptical isophotes of axis ratio~$q$ and orientation~$\varphi$, the most important ellipticity descriptor for weak lensing (hereafter simply \emph{the ellipticity}),
\begin{equation}\label{eq:eps-q}
	\epsilon = \frac{1 - q}{1 + q} \, \E^{2 \, \I \, \varphi} \;,
\end{equation}
was introduced by \citet{1997A&A...318..687S} because it provides an unbiased estimator for the shear of weak gravitational lensing:
The (reduced) shear~$g$ acts on the observed ellipticity~$\epsilon$,
\begin{equation}
	\epsilon = \frac{\epsilon_s + g}{1 + g^* \epsilon_s} \;,
\end{equation}
where $\epsilon_s$ is the source ellipticity before lensing.
For an isotropic source distribution with uniformly distributed direction~$\arg \epsilon_s$, the observed ellipticity recovers the shear,
\begin{equation}
	\Ev[\epsilon] = g \;.
\end{equation}
Hence, by observing the ellipticities of many astronomical sources, we can infer the properties of the shear field.
This is the basic idea behind cosmology with cosmic shear surveys (for a recent review, see \citealp{2015RPPh...78h6901K}).

If the morphology is more complicated than a simple ellipse, the shape of a source must be defined in terms of its intrinsic central image moments~$\mu_{pq}$,
\begin{equation}
    \mu_{pq} = \int \! I(x, y) \, (x - \bar{x})^p \, (y - \bar{y})^q \, \D x \, \D y \;,
\end{equation}
where~$I(x, y)$ is the surface brightness distribution of the source, and $(\bar{x}, \bar{y})$ is its centroid.
The ellipticity~$\epsilon$ can then be expressed as a particular combination of the image moments,
\begin{equation}\label{eq:eps-mu}
    \epsilon = \frac{\mu_{20} - \mu_{02} + 2 \, \I \, \mu_{11}}{\mu_{20} + \mu_{02} + 2 \sqrt{\mu_{20} \mu_{02} - \mu_{11}^2}} \;.
\end{equation}
It is convenient to introduce the \emph{Stokes parameters} $u, v, s$ for the individual terms in this relation \citep*{2014MNRAS.439.1909V},%
\begin{equation}\label{eq:uvs}
    u = \mu_{20} - \mu_{02} \;, \quad
    v = 2 \mu_{11} \;, \quad
    s = \mu_{20} + \mu_{02} \;,
\end{equation}
so that the ellipticity can be expressed equivalently and more easily in terms of the Stokes parameters,
\begin{equation}\label{eq:eps}
    \epsilon = \frac{u + \I \, v}{s + \sqrt{s^2 - u^2 - v^2}} \;,
\end{equation}
where $u^2 + v^2 < s^2$ is guaranteed by the Cauchy-Schwarz inequality for the second-order moments.

Most shear measurement methods thus ultimately fall into one of two categories:
either fit an analytical elliptical galaxy profile to the observed image to obtain the ellipticity~\eqref{eq:eps-q} from the model, or measure the moments of the image to calculate the ellipticity~\eqref{eq:eps-mu} directly.

In the following, we present a new moment-based approach to shape measurement.
However, instead of treating it as a problem of computer science or image analysis, we will try to reduce ellipticity estimation to its most basic statistical form.
In Section~\ref{sec:moments}, we use unweighted image moments to recover the Stokes parameters~$u, v, s$ from noisy observations as the means of three independent normal random variables with known variance.
Measuring the ellipticity~\eqref{eq:eps} then becomes a problem of statistical parameter estimation, and in Section~\ref{sec:estimation} we derive some of the general results this approach allows us to make.
As a practical application, Section~\ref{sec:ube} contains the derivation of an unbiased ellipticity estimator.
Finally, we discuss the results in Section~\ref{sec:discussion} and present our conclusions in Section~\ref{sec:conclusion}.

\section{Distribution of the Moments}
\label{sec:moments}

To measure the shapes of sources, we expand on earlier results \citep{2017MNRAS.471L..57T} to determine the second-order image moments of the source, in the form of the Stokes parameters~$u, v, s$, so that we may compute the ellipticity~\eqref{eq:eps}.
The main problem is the pixel noise that overlays the observations, which turns the measurement process into an exercise in statistics, so that we can only infer the true value of the parameters from the distribution of the moments we observe.
However, even before we deal with the statistical properties of the noise, we must consider the effects of the observational process on the observed signal itself.

\subsection{Signal}
\label{sec:signal}

An observed source has undergone a number of effects that influence its signal.
The influence of imperfect optics and seeing is modelled by convolution with a \emph{point-spread function} (PSF).
The observed signal with PSF~$P$ is
\begin{equation}
    I^{P}(x,y)
    = \int \! P(x-x', y-y') \, I(x',y') \, \D x' \, \D y' \;,
\end{equation}
or, in short, $I^{P} = P*I$.
The signal is subsequently collected in pixels of a finite resolution, and the observed signal~$I^{PD}_k$ in pixel~$k$ is discretised,
\begin{equation}\label{eq:I_PD_k}
    I^{PD}_k = \int_{A_k} \! I^{P}(x,y) \, \D x \, \D y \;,
\end{equation}
where $A_k$ is the area of the pixel.
If all pixels have the same area~$A$, we can define the discretisation kernel~$D$ as the normalised window function of the pixels,
\begin{equation}
    D(x,y) \equiv \frac{1}{|A|} \; \begin{cases}
        \, 1 & \text{if $(x,y) \in A$,} \\
        \, 0 & \text{if $(x,y) \notin A$,}
    \end{cases}
\end{equation}
so that the discretisation can be carried out as another convolution,
\begin{equation}\label{eq:I_PD}
    I^{PD}(x, y)
    = \int \! D(x-x', y-y) \, I^{P}(x', y') \, \D x' \, \D y' \;,
\end{equation}
or $I^{PD} = D*I^{P} = D*P*I$.
This last convolution~\eqref{eq:I_PD} then recovers the discretised signal~\eqref{eq:I_PD_k},
\begin{equation}\label{eq:I_PD_discretise}
    I^{PD}_k = A_k \, I^{PD}(x_k, y_k) \;,
\end{equation}
when evaluated at the location $(x_k, y_k)$ of pixel~$k$.

We can now compute central moments~$\mu_{pq}^{PD}$ for the observed signal $I^{PD}$ with PSF and discretisation,
\begin{equation}\label{eq:mu_PD_pq}
    \mu^{PD}_{pq}
    = \int \! I^{PD}(x, y) \, (x - \tilde x)^p \, (y - \tilde y)^q \, \D x \, \D y \;,
\end{equation}
where $(\tilde x, \tilde y)$ is the centroid after convolution.
Since the observed signal is discretely sampled by the pixel grid, the integral~\eqref{eq:mu_PD_pq} must be approximated by a sum,
\begin{align}\label{eq:mu_PD_pq-approx}
    \mu^{PD}_{pq}
    &\approx \sum_{k} w_k \, I^{PD}(x_k, y_k) \, (x_k - \tilde x)^p \, (y_k - \tilde y)^q \, A_k \nonumber \\
    &= \sum_{k} w_k \, I^{PD}_k \, (x_k - \tilde x)^p \, (y_k - \tilde y)^q \;,
\end{align}
where the relation~\eqref{eq:I_PD_discretise} between the continuous and discrete versions of signal~$I^{PD}$ has been used.
Here, we have introduced a mask $w_k$ that selects which pixels contribute to the integral, and it is assumed that $w_k \equiv 1$ wherever the signal of the source does not vanish.\footnote{%
The mask is not used to give individual weights to pixels, but merely as a tool to describe the aperture in which pixels are considered.}

Finally, we wish to relate the moments of the observed signal to the intrinsic moments of the source.
For this, we can use the relation between the moments of a convolution $I^{PD} = D*P*I$ and its constituent functions (see Appendix~\ref{sec:moments-convolution} for a short derivation).
Denoting the central moments of the functions~$P$ and~$D$ with~$\pi_{pq}$ and~$\delta_{pq}$, respectively, the second-order central moments of the convolution are
\begin{equation}\label{eq:mu_PD_pq-rel}
    \mu^{PD}_{pq}
    = \mu_{pq} + \mu_{00} \, (\pi_{pq} + \delta_{pq}) \;,
    \quad p + q = 2 \;,
\end{equation}
where both the PSF and discretisation kernels are normalised with moments $\pi_{00} = \delta_{00} = 1$.
Convolution with a normalised kernel does not change the total signal, $\mu^{PD}_{00} = \mu^{}_{00}$, so that we can rearrange the relation~\eqref{eq:mu_PD_pq-rel} for the intrinsic second-order central moments,
\begin{equation}\label{eq:mu_pq-rel}
    \mu_{pq} =
    \mu^{PD}_{pq} - \mu^{PD}_{00} \, (\pi_{pq} + \delta_{pq}) \;,
    \quad p + q = 2 \;,
\end{equation}
in terms of the central moments of the observed signal, PSF, and discretisation kernel, respectively.
Inserting the discrete form~\eqref{eq:mu_PD_pq-approx} of the moments~$\mu^{PD}$ into relation~\eqref{eq:mu_pq-rel} then yields an expression for the intrinsic second-order central moments directly in terms of the observed signal available in the pixels,
\begin{equation}\label{eq:mu-pq}
    \mu_{pq}
    \approx \sum_{k} w_k \, I^{PD}_k \, \big[(x_k - \tilde x)^p \, (y_k - \tilde y)^q - \nu_{pq}\big] \;,
\end{equation}
where $p + q = 2$ as before, and $\nu_{pq} = \pi_{pq} + \delta_{pq}$ is the sum of the moments of the convolution kernels.

Once the moments~\eqref{eq:mu-pq} are obtained, the definition~\eqref{eq:uvs} yields the intrinsic Stokes parameters~$u, v, s$ of the source from the observed signal,
\begin{align}
    u &\approx \sum_{k} w_k \, I^{PD}_k \, \big[(x_k - \tilde x)^2  - (y_k - \tilde y)^2 - \nu_{20} + \nu_{02}\big] \;, \label{eq:u-sig} \\
    v &\approx \sum_{k} w_k \, I^{PD}_k \, \big[2 \, (x_k - \tilde x) \, (y_k - \tilde y)^2 - 2 \, \nu_{11}\big] \;, \label{eq:v-sig} \\
    s &\approx \sum_{k} w_k \, I^{PD}_k \, \big[(x_k - \tilde x)^2  + (y_k - \tilde y)^2 - \nu_{20} - \nu_{02}\big] \label{eq:s-sig} \;.
\end{align}
The approximation is due to the discretisation~\eqref{eq:mu_PD_pq-approx}, and how well it works in practice depends on the relative size of the source and resolution.
In the following, we will assume that equality holds for the Stokes parameters~\eqref{eq:u-sig}--\eqref{eq:s-sig}, and that the total moments~$\nu_{pq}$ of PSF and pixellation are known.

\subsection{Noise}

In any real observation, the signal is effectively overlaid with noise.
For the purpose of this derivation, we assume that the noise in the observed pixels is normally distributed with zero mean, but not necessarily uncorrelated.
Especially for ground-based observations, normality of the noise is usually a very good assumption, since the main contributor is Poisson noise from atmospheric emission at a high background count level.

The pixel data is thus assumed to be a multivariate normal random vector~$\vec{D}$ with the mean in pixel~$k$ given by the signal~$I^{PD}_k$ defined above,
\begin{equation}
    \Ev[D_k] = I^{PD}_k \;,
\end{equation}
and a covariance matrix~$\mat\Sigma$ describing the covariance between any two pixels~$i$ and~$j$,
\begin{equation}
    \Cov[D_i, D_j] = \Sigma_{ij} \;.
\end{equation}
Here and in the following, the covariance matrix for the pixel noise is assumed known.

The quantities of interest for shape description are the Stokes parameters $u,v,s$.
Above, it was shown how these can be expressed as the linear combinations~\eqref{eq:u-sig}--\eqref{eq:s-sig} of the signal in each pixel.
We therefore define the random variables $X, Y, Z$ in the same way using the random pixel data vector~$\vec{D}$,
\begin{align}
    X &= \sum_{k} w_k \, D_k \, \big[(x_k - \tilde x)^2  - (y_k - \tilde y)^2 - \nu_{20} + \nu_{02}\big] \;, \label{eq:X} \\
    Y &= \sum_{k} w_k \, D_k \, \big[2 \, (x_k - \tilde x) \, (y_k - \tilde y)^2 - 2 \, \nu_{11}\big] \;, \label{eq:Y} \\
    Z &= \sum_{k} w_k \, D_k \, \big[(x_k - \tilde x)^2  + (y_k - \tilde y)^2 - \nu_{20} - \nu_{02}\big] \label{eq:Z} \;.
\end{align}
The expectation of $X, Y, Z$ then recovers, within the approximations made above, the Stokes parameters,
\begin{equation}\label{eq:XYZ-mean}
    \Ev[X] = u \;, \quad
    \Ev[Y] = v \;, \quad
    \Ev[Z] = s \;.
\end{equation}
The linear relations~\eqref{eq:X}--\eqref{eq:Z} between the random vectors~$\vec{D}$ and~$(X, Y, Z)$ can be written more concisely in terms of a matrix~$\mat{M}$ with three rows and columns for each pixel~$k$,
\begin{align}
    M_{1k} &= w_k \, \bigl[(x_k - \tilde x)^2  - (y_k - \tilde y)^2 - \nu_{20} + \nu_{02}\bigr] \;, \label{eq:M_1k} \\
    M_{2k} &= w_k \, \bigl[2 \, (x_k - \tilde x) \, (y_k - \tilde y)^2 - 2 \, \nu_{11}\bigr] \;, \label{eq:M_2k} \\
    M_{3k} &= w_k \, \bigl[(x_k - \tilde x)^2  + (y_k - \tilde y)^2 - \nu_{20} - \nu_{02}\bigr] \label{eq:M_3k} \;,
\end{align}
so that $(X, Y, Z) = \mat{M} \, \vec{D}$.
Since $\vec{D}$ is a multivariate normal random vector, this implies that $(X, Y, Z)$ is multivariate normal as well, with its mean given by the expectation~\eqref{eq:XYZ-mean},
\begin{equation}
    \Ev[(X, Y, Z)] = (u, v, s) \;,
\end{equation}
and the covariance matrix~$\mat{C} = \Cov[(X, Y, Z)]$ as usual for linear transformations,
\begin{equation}\label{eq:C}
    \mat{C} = \mat{M} \, \mat{\Sigma} \, \mat{M}^\tp \;,
\end{equation}
with entries $C_{ij} = \sum_{kl} M_{ik} M_{jl} \, \Sigma_{kl}$.
Both matrices~$\mat{M}$ and~$\mat{\Sigma}$ are known, so that we can compute the covariance matrix~$\mat{C}$ of the random variables~$X, Y, Z$ without problem.

\subsection{Multiple Exposures}

Multiple exposures of the same intrinsic signal can be combined directly at the level of the random variables $X, Y, Z$ without the need for co-addition of the individual observations.

Let $X_i, Y_i, Z_i$ be the jointly normal random variables~\eqref{eq:X}--\eqref{eq:Z} obtained for the $i$'th exposure individually, with the means~\eqref{eq:XYZ-mean} replaced by~$u_i, v_i, s_i$ and the covariance matrix~\eqref{eq:C} replaced by~$\mat{C}_i$.
Fixing an reference frame in which the Stokes parameters are $u, v, s$, the parameters $u_i, v_i, s_i$ of the $i$'th exposure are related by the rotation angle~$\alpha_i$ between the coordinate systems,
\begin{equation}
    \begin{pmatrix}
        u \\ v \\ s
    \end{pmatrix}
    = \begin{pmatrix}
        \cos 2 \alpha_i & -\sin 2 \alpha_i & 0 \\
        \sin 2 \alpha_i &  \cos 2 \alpha_i & 0 \\
                      0 &                0 & 1
    \end{pmatrix} \, \begin{pmatrix}
        u_i \\ v_i \\ s_i
    \end{pmatrix}
    \equiv \mat{R}_i \, \begin{pmatrix}
        u_i \\ v_i \\ s_i
    \end{pmatrix} \;,
\end{equation}
where $\mat{R}_i$ is the rotation matrix for exposure~$i$.
Due to linearity, the transformed vectors $\mat{R}_i \, (X_i, Y_i, Z_i)$ remain multivariate normal, now with the same means $\Ev[\mat{R}_i \, (X_i, Y_i, Z_i)] = (u, v, s)$ but potentially different covariance matrices $\Cov[\mat{R}_i \, (X_i, Y_i, Z_i)] = \mat{R}_i \, \mat{C}_i \, \mat{R}_i^\tp$.

\begin{figure*}%
\centering%
\includegraphics[scale=0.66]{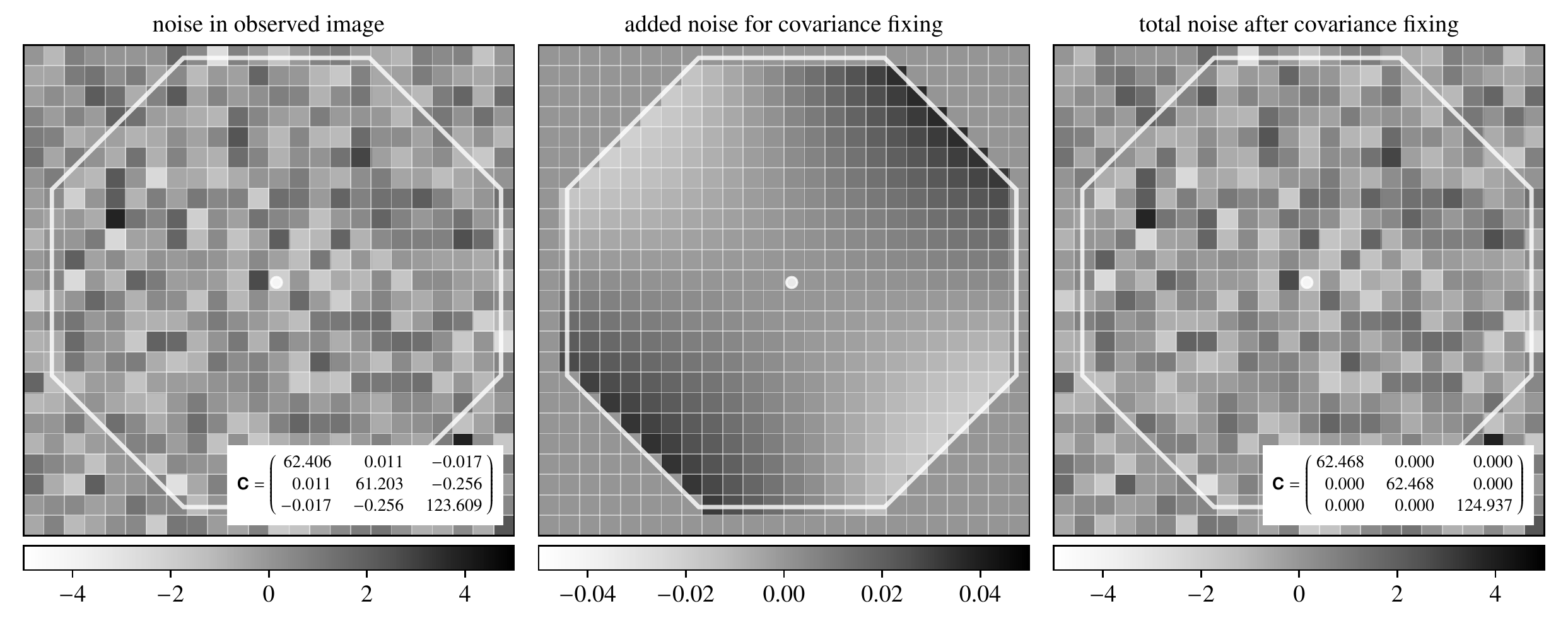}%
\caption{%
Covariance fixing for an observed image with the indicated centroid and mask (\emph{white}).
\emph{Left:}
Uncorrelated noise in the observed image.
The covariance matrix~$\mat{C}$ is close to diagonal with approximate ratios $1 : 1 : 2$.
\emph{Centre:}
Correlated noise added to the image, generated from three independent normal variates as described in Appendix~\ref{sec:covariance-fixing}.
\emph{Right:}
Total noise after covariance fixing.
Since the magnitude of the added noise is about 100 times lower, the overall noise level has not changed.
The fixed covariance matrix~$\mat{C}$ is now exactly diagonal with ratios $1 : 1 : 2$.
}%
\label{fig:cfix}%
\end{figure*}

These rotated random vectors are hence multiple independent observations of the Stokes parameters~$u, v, s$ and can be combined as a vector-valued weighted mean, where the individual weight matrix~$\mat{W}_i$ for exposure~$i$ is the inverse of the covariance matrix,
\begin{equation}
    \mat{W}_i = \bigl(\mat{R}_i \, \mat{C}_i \, \mat{R}_i^\tp\bigr)^{-1} \;.
\end{equation}
The combined random vector $(X, Y, Z)$ is a linear combination of independent normal random vectors,
\begin{equation}
    (X, Y, Z)
    = \Bigl(\sum_i \mat{W}_i\Bigr)^{-1} \sum_i \mat{W}_i \, \mat{R}_i \, (X_i, Y_i, Z_i)\;,
\end{equation}
and remains multivariate normal with mean $\Ev[(X, Y, Z)] = (u, v, s)$ and covariance matrix~$\mat{C} = \Cov[(X, Y, Z)]$ given by the inverse sum of the weight matrices,
\begin{equation}\label{eq:C-mult}
    \mat{C} = \Bigl(\sum_i \mat{W}_i\Bigr)^{-1} \;.
\end{equation}
In the special case of identical covariance matrices~$\mat{C}_i = \mat{C}_1$ and no rotation, the combined covariance matrix~$\mat{C}$ for $N$ individual exposures is then $\mat{C} = \mat{C}_1/N$ as expected.

\subsection{Covariance Matrix}

We now show that the covariance matrix~$\mat{C}$ for the combinations of moments $X, Y, Z$ is often approximately diagonal.
Subsequently, we also show how this can generally and exactly be achieved by a simple manipulation of the data.

Under a rotation of $45\degr$, the coordinates in the first two rows~\eqref{eq:M_1k} and~\eqref{eq:M_2k} of matrix~$\mat{M}$ transform as $x^2 - y^2 \mapsto 2 x y$ and $2 xy \mapsto -x^2 + y^2$, whereas the coordinates in the third row~\eqref{eq:M_3k} remain unchanged, $x^2 + y^2 \mapsto x^2 + y^2$.
Neglecting the $\nu_{pq}$-terms for the moment, the matrix~$\mat{M}'$ after a rotation of $45\degr$ about the centroid~$(\tilde x, \tilde y)$ is thus approximately
\begin{equation}
    M'_{1k} \approx M_{2k} \;, \quad
    M'_{2k} \approx -M_{1k} \;, \quad
    M'_{3k} = M_{3k} \;.
\end{equation}
This transformation of the matrix~$\mat{M}$ is also reflected in the entries of the covariance matrix~\eqref{eq:C},
\begin{equation}
    \mat{C}' \approx \begin{pmatrix*}[r]
         C_{22} & -C_{12} &  C_{23} \\
        -C_{12} &  C_{11} & -C_{13} \\
         C_{23} & -C_{13} &  C_{33}
    \end{pmatrix*} \;,
\end{equation}
where $C_{ij}$ are the entries of the original covariance matrix~$\mat{C}$.
Under a further rotation of $45\degr$ the transformation is similar,
\begin{equation}
    \mat{C}'' \approx \begin{pmatrix*}[r]
         C'_{22} & -C'_{12} &  C'_{23} \\
        -C'_{12} &  C'_{11} & -C'_{13} \\
         C'_{23} & -C'_{13} &  C'_{33}
    \end{pmatrix*} \approx \begin{pmatrix*}[r]
         C_{11} &  C_{12} & -C_{13} \\
         C_{12} &  C_{22} & -C_{23} \\
        -C_{13} & -C_{23} &  C_{33}
    \end{pmatrix*} \;.
\end{equation}
If the mask~$w_k$ and pixel covariance matrix~$\mat{\Sigma}$ are approximately invariant under $45\degr$ rotations, then $\mat{C} \approx \mat{C}' \approx \mat{C}''$, which implies that the off-diagonals approximately vanish, $C_{ij} \approx 0$ for $i \neq j$.
Furthermore, we also see that in this case, the first two diagonal entries are approximately equal, $C_{11} \approx C_{22}$.

Additional progress can be made for uncorrelated pixel noise, in which case the covariance matrix~$\mat{\Sigma}$ is diagonal, $\Sigma_{ij} = \sigma_i^2 \, \delta_{ij}$, where $\delta_{ij}$ is the Kronecker delta.
The diagonal of the covariance matrix~$\mat{C}$ then simplifies,
\begin{equation}
    C_{ii}
    = \sum_{kl} M_{ik} M_{jl} \, \sigma_k^2 \, \delta_{kl}
    = \sum_{k} M_{ik}^2 \, \sigma_k^2 \;,
\end{equation}
and since $(x^2 - y^2)^2 + (2xy)^2 = (x^2 + y^2)^2$, we have the further approximations $M_{1k}^2 + M_{2k}^2 \approx M_{3k}^2$ and $C_{11} + C_{22} \approx C_{33}$.

Combining all of these relations, we find that the covariance matrix~$\mat{C}$ of $(X, Y, Z)$ is approximately diagonal,
\begin{equation}\label{eq:C-diag}
    \mat{C} \approx \begin{pmatrix}
        \sigma_X^2 & 0          & 0          \\
        0          & \sigma_Y^2 & 0          \\
        0          & 0          & \sigma_Z^2
    \end{pmatrix} \;,
\end{equation}
with $\sigma_X^2 \approx \sigma_Y^2 \approx \frac{1}{2} \sigma_Z^2$ the variances of $X, Y, Z$, respectively.
This holds, provided that the mask, pixel noise covariance matrix, and pixel grid are all invariant under rotations of $45\degr$, that the pixel noise is uncorrelated, and that the correction terms $\nu_{pq}$ due to PSF and pixellation are negligible.
When multiple exposures are combined, each individual covariance matrix in the sum~\eqref{eq:C-mult} remains of the form~\eqref{eq:C-diag} even after rotation, and the combined covariance matrix is approximately diagonal also in this case.

It is of course impossible to fulfil these assumptions exactly.
However, given uncorrelated and homoscedastic pixel noise, and choosing a 45\degr-symmetric (e.g.\@ octagonal or circular) mask, the covariance matrix is typically very close to the approximate form.
Furthermore, in Appendix~\ref{sec:covariance-fixing} we present a simple and efficient recipe to fix the covariance matrix~$\mat{C}$ to any desired shape \emph{exactly}, by adding a small amount of noise to the image (Fig.~\ref{fig:cfix}).
Therefore, we can generally assume that the covariance matrix~$\mat{C}$ is diagonal with entries $\sigma_X^2 = \sigma_Y^2 = \sigma^2$, $\sigma_Z^2 = 2 \sigma^2$, for some variance~$\sigma^2$.

\section{Ellipticity Estimation}
\label{sec:estimation}

We have seen that it is possible to reduce the observed data into the jointly normal and independent random variables~$X, Y, Z$ of fixed variance~$\sigma^2, \sigma^2, 2\sigma^2$ that recover the Stokes parameters~$u, v, s$ through their means.
Since, the ellipticity~\eqref{eq:eps} is directly related to these means, we can make a number of general observations about how the parameter~$\epsilon$ can be estimated.
We first find the relevant signal-to-noise ratio for ellipticity estimation with this method.
We then set out a generic form for an estimator.
Finally, we calculate the lower bound on the variance of an unbiased ellipticity estimator.

\subsection{Signal-to-Noise Ratio}

How much we can trust an estimate of the ellipticity must naturally depend on the amount of signal and noise in the data.
To understand this dependency, we can rewrite the Stokes parameters~$u$ and~$v$ in terms of~$s$ and~$\epsilon$ using the definition~\eqref{eq:eps} of ellipticity,
\begin{equation}
	u + \I \, v
	= \frac{2 \, s \, \epsilon}{1 + |\epsilon|^2} \;,
\end{equation}
and use this relation to express the random variables~$X, Y, Z$ through independent standard normal random variables~$\tilde{X}, \tilde{Y}, \tilde{Z}$,
\begin{align}
	X &= s \left( \frac{2 \Re \epsilon}{1 + |\epsilon|^2} + \frac{\sigma}{s} \, \tilde{X}\right) \;, \\
	Y &= s \left( \frac{2 \Im \epsilon}{1 + |\epsilon|^2} + \frac{\sigma}{s} \, \tilde{Y}\right) \;, \\
	Z &= s \left(1 + \sqrt{2} \, \frac{\sigma}{s} \, \tilde{Z} \right) \;.
\end{align}
Apart from an overall scaling, we find that the distributions depend only on the ellipticity~$\epsilon$ and the quotient~$\sigma/s$.
Its inverse,
\begin{equation}\label{eq:snr}
    \SNR = \frac{s}{\sigma} \;,
\end{equation}
is the relevant signal-to-noise ratio for ellipticity estimation with unweighted image moments.

In observations, the true value of~$s$, and consequently $\SNR$, is of course unknown.
However, it can be straightforwardly estimated for an individual object using the random variable~$Z$,
\begin{equation}
    \ESNR = \frac{Z}{\sigma} \;,
\end{equation}
with expectation~$\Ev[\ESNR] = \SNR$ and variance~$\Var[\ESNR] = 2$.
For a survey of a population of objects, the distribution of estimated signal-to-noise ratios~$\ESNR_i$ has the mean~$\Ev[\ESNR_i] = \mu_\SNR$ and variance~$\Var[\ESNR_i] = \sigma_\SNR^2 + 2$, where $\mu_\SNR$ and $\sigma_\SNR^2$ are the mean and variance of the population, respectively.
In this way, we can recover the signal-to-noise properties of the survey from the distribution of the estimates.

\subsection{Generic Form of the Ellipticity Estimator}
\label{sec:estim}

\begin{figure}%
\centering%
\includegraphics[scale=0.66]{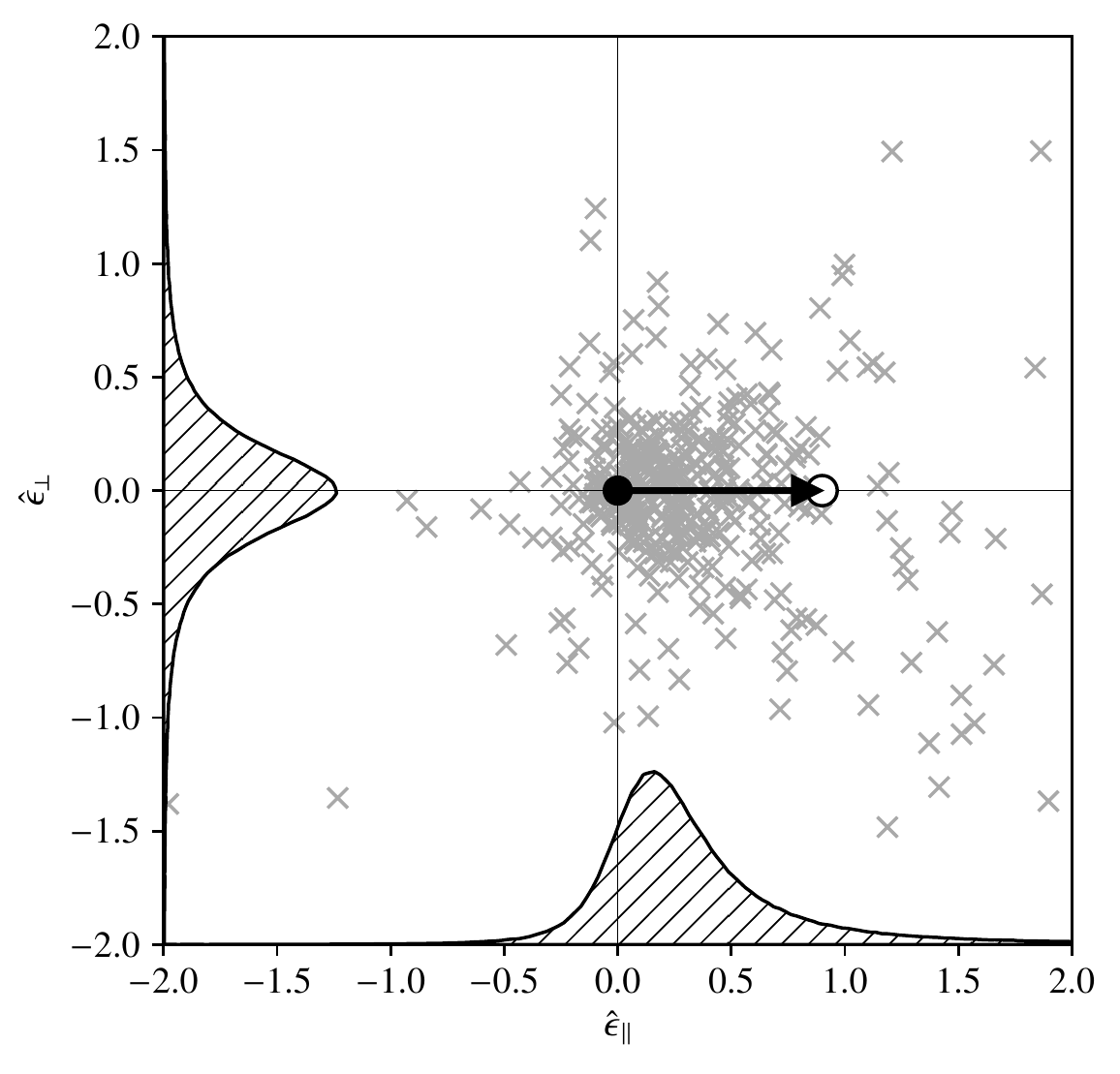}%
\caption{%
Marginal distributions for the two components of the ellipticity estimates (\emph{crosses}).
The components~$\hat{\epsilon}_{\,\parallel}$ and~$\hat{\epsilon}_{\perp}$ are defined parallel and perpendicular to the direction (\emph{black arrow}) of the true ellipticity (\emph{open circle}).
The marginal distribution of~$\hat{\epsilon}_{\perp}$ is symmetric about the origin for any estimator of the form~\eqref{eq:estim}.
}%
\label{fig:comp}%
\end{figure}

It turns out that there is a useful generic form for an ellipticity estimator~$\hat{\epsilon}$ using the combinations of moments~$X, Y, Z$,
\begin{equation}\label{eq:estim}
    \hat{\epsilon}
    = (X + \I \, Y) \, h\big(\sqrt{X^2 + Y^2}, Z\big) \;,
\end{equation}
where $h$ is a free real-valued function of $\sqrt{X^2 + Y^2}$ and $Z$ only.
We can write the expectation of such an estimator,
\begin{equation}
   \Ev[\hat{\epsilon}]
   = \int \! \frac{r \, \E^{\I \, \varphi} \, h(r, z)}{4 \pi^{3/2} \sigma^3} \, \E^{-\frac{r^2 + t^2 - 2 r t \cos(\varphi - \vartheta)}{2 \sigma^2} - \frac{(z - s)^2}{4\sigma^2}} \, r \, \D r \, \D \varphi \, \D z \;,
\end{equation}
where the polar coordinates $x + \I \, y = r \, \E^{\I \varphi}$ and $u + \I \, v = t \, \E^{\I \vartheta}$ have been introduced.
Using an integral representation of the modified Bessel function \citep[8.431.5]{2007tisp.book.....G},
\begin{equation}
   I_1(x)
   = \frac{1}{\pi} \int_{0}^{\pi} \! \E^{x \cos \varphi} \cos \varphi \, \D \varphi
   = \frac{1}{2\pi} \int_{-\pi}^{\pi} \! \E^{\I \, \varphi + x \cos \varphi} \, \D \varphi \;,
\end{equation}
the angular integration can be carried out after a translation by~$\vartheta$.
The resulting expectation,
\begin{equation}\label{eq:e-eps}
   \Ev[\hat{\epsilon}]
   = \E^{\I \, \vartheta} \int \! \frac{r \, h(r, z)}{2 \sqrt{\pi} \sigma^3} \, \E^{-\frac{2 r^2 + 2 t^2 + (z - s)^2}{4\sigma^2}} \, I_1\Big(\frac{r t}{\sigma^2}\Big) \, r \, \D r \, \D z \;,
\end{equation}
shows that an estimator of the form~\eqref{eq:estim} produces an estimate that is unbiased in the ellipticity angle~$\vartheta$, independent of the choice of function~$h(r, z)$.

In the framework presented here, the performance of a given ellipticity estimator can be analysed directly by evaluating the estimates for sets of independent normal random variates~$X, Y, Z$ with respective means $u, v, s$ and variances~$\sigma^2, \sigma^2, 2\sigma^2$, which are easily generated.
For an estimator of the generic form~\eqref{eq:estim}, it is useful to further decompose the estimates into a parallel component~$\hat{\epsilon}_{\,\parallel}$ and a perpendicular component~$\hat{\epsilon}_{\perp}$ relative to the direction of the true ellipticity~$\epsilon$ (Fig.~\ref{fig:comp}).
The distribution of the perpendicular component~$\hat{\epsilon}_{\perp}$ is always symmetric about the origin, and the estimator is unbiased in the ellipticity angle.
The overall performance of the estimator is therefore mostly dependent on the distribution of the parallel component~$\hat{\epsilon}_{\,\parallel}$.

\subsection{Cram\'er-Rao Lower Bounds}

\begin{figure}%
\centering%
\includegraphics[scale=0.66]{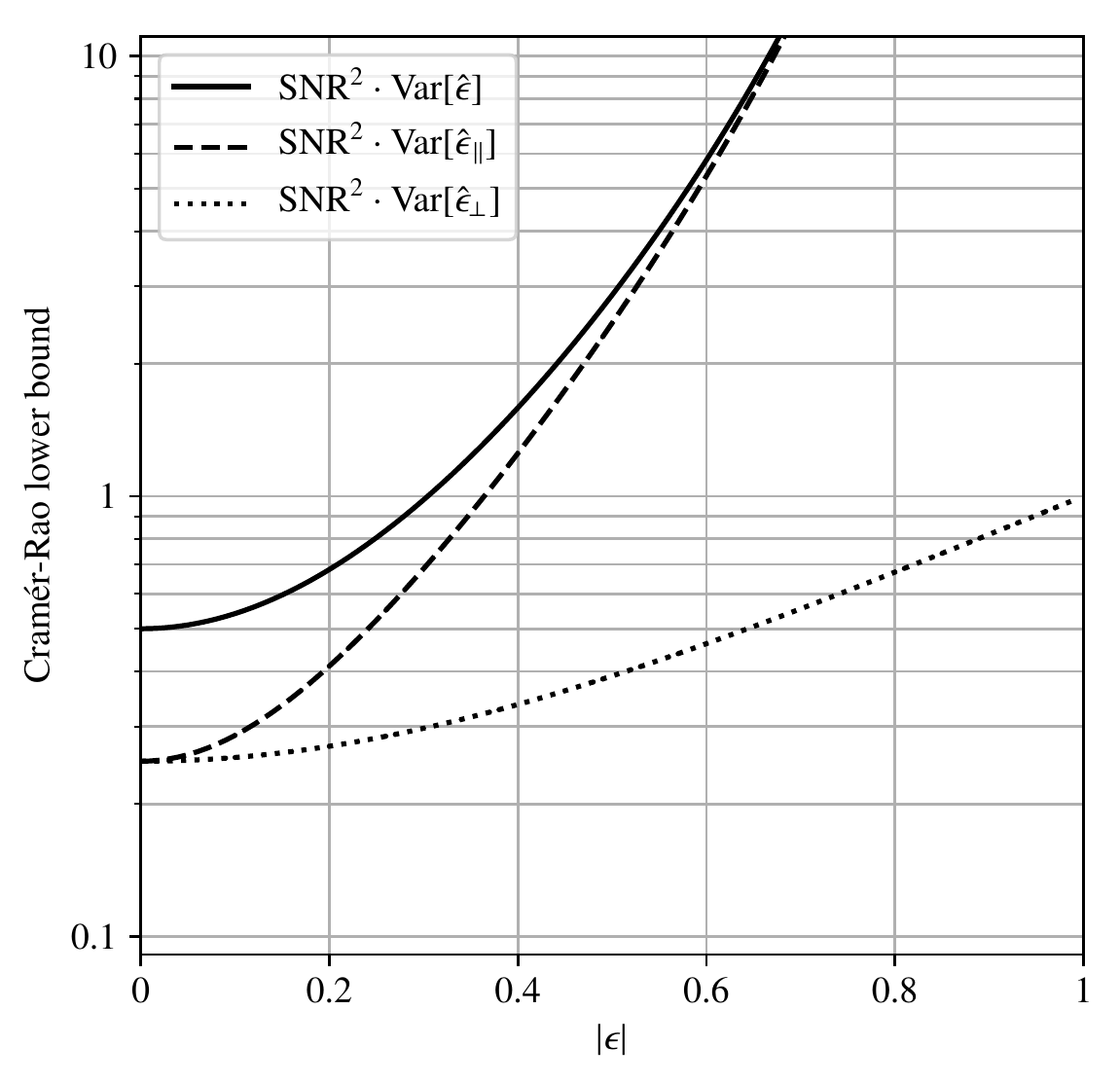}%
\caption{%
The Cram\'er-Rao lower bounds for the variance of an unbiased estimator for the complex ellipticity (\emph{solid}) and its individual components parallel (\emph{dashed}) and perpendicular (\emph{dotted}) to the true ellipticity.
The curves are drawn as functions of true ellipticity modulus~$|\epsilon|$ and normalised by the square of the signal-to-noise ratio~$\SNR$.
}%
\label{fig:crlb}%
\end{figure}

As the random variables $X, Y, Z$ are jointly normal, and their means~$u, v, s$ are related to the ellipticity~$\epsilon$ through definition~\eqref{eq:eps}, it is possible to compute the Cram\'er-Rao lower bound (CRLB) for the variance of an ellipticity estimator.
For an unbiased estimator of the complex parameter~$\epsilon$ \citep{1994ITSP...42.2859V},
\begin{equation}\label{eq:crlb}
	\Var[\hat{\epsilon}]
	\ge \frac{(1 + |\epsilon|^2)^2 \, (1 + 4 |\epsilon|^2 + |\epsilon|^4)}{2 \, (1 - |\epsilon|^2)^2} \, \frac{\sigma^2}{s^2} \;,
\end{equation}
where $\Var[\hat{\epsilon}] = \Ev[|\hat{\epsilon} - \epsilon|^2]$ is the absolute variance of a complex-valued estimator.
The CRLB is inversely proportional to the square of the signal-to-noise ratio~\eqref{eq:snr}, and diverges as the ellipticity modulus~$|\epsilon|$ approaches unity (Fig.~\ref{fig:crlb}).

Using the parallel and perpendicular ellipticity components introduced in the previous section, the CRLB can be computed for unbiased estimators of the individual components~$\epsilon_{\,\parallel}$ and~$\epsilon_{\perp}$,
\begin{gather}
	\Var[\hat{\epsilon}_{\,\parallel}]
	\ge \frac{(1 + |\epsilon|^2)^2 \, (1 + 10 |\epsilon|^2 + |\epsilon|^4)}{4 \, (1 - |\epsilon|^2)^2} \, \frac{\sigma^2}{s^2} \;, \label{eq:crlb-par} \\
	\Var[\hat{\epsilon}_{\perp}]
	\ge \frac{(1 + |\epsilon|^2)^2}{4} \, \frac{\sigma^2}{s^2} \;, \label{eq:crlb-perp}
\end{gather}
where the individual bounds for the components sum to the bound~\eqref{eq:crlb} for the total variance.
This is particularly interesting since every estimator of the generic form~\eqref{eq:estim} is unbiased in the perpendicular component~$\hat{\epsilon}$, for which the CRLB does not diverge as the ellipticity approaches unity.

\section{The Unbiased Ellipticity Estimator}
\label{sec:ube}

To demonstrate the power of a full statistical description of the ellipticity estimation problem, in the form of the independent normal random variables~$X, Y, Z$ with the Stokes parameters~$u, v, s$ as their means, we now derive an unbiased estimator for the ellipticity~\eqref{eq:eps}.

\subsection{Derivation of the Unbiased Estimator}

\begin{figure}%
\centering%
\includegraphics[scale=0.66]{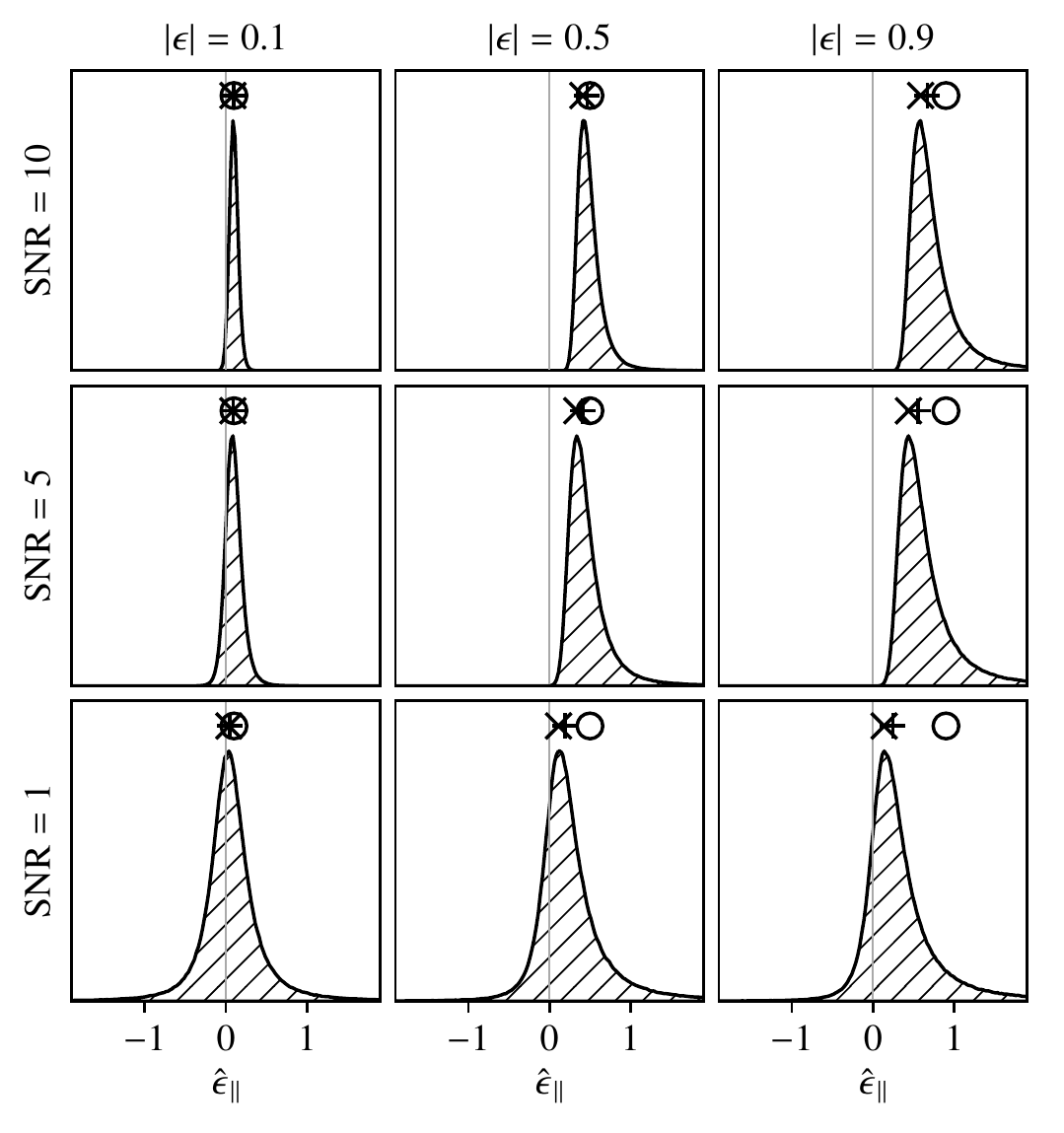}%
\caption{%
The distribution of the parallel component~$\hat{\epsilon}_{\,\parallel}$ of the unbiased ellipticity estimator, with mode (\emph{cross}), median (\emph{plus}), and mean (\emph{circle}).
For high ellipticity magnitudes~$|\epsilon|$, the distribution increasingly skews to the right.
For low signal-to-noise ratios~$\SNR$, the mode of the distribution goes towards zero ellipticity.
}%
\label{fig:dist}%
\end{figure}

We have seen above that there is a class of estimators that are unbiased in the ellipticity angle.
We now go one step further and \emph{require} the estimator to produce an unbiased estimate of the full ellipticity~$\epsilon$,%
\begin{equation}
    \Ev[\hat{\epsilon}]
    \overset{!}{=} \epsilon
    = \frac{t \, \E^{\I \, \vartheta}}{s + \sqrt{s^2 - t^2}} \;,
\end{equation}
where the second equality is the ellipticity definition~\eqref{eq:eps} in polar coordinates~$u + \I \, v = t \, \E^{\I \vartheta}$.
For an estimator of the generic form~\eqref{eq:estim}, we have already found that the expectation~\eqref{eq:e-eps} on the left-hand side contains the requisite factor of $\E^{\I \, \vartheta}$, so that the remaining terms can be rearranged into an integral equation for the free function~$h$,
\begin{equation}\label{eq:h-eq}
    \int \! \frac{r \, h(r, z)}{2 \sqrt{\pi} \sigma^3} \, \E^{-\frac{2 r^2 + z^2}{4 \sigma^2}} \, \E^{\frac{z s}{2 \sigma^2}} \, I_1\Big(\frac{r t}{\sigma^2}\Big) \, r \, \D r \, \D z
    = \frac{t \, \E^{\frac{2 t^2 + s^2}{4\sigma^2}}}{s + \sqrt{s^2 - t^2}} \;.
\end{equation}
The algebraic factor on the right-hand side can be written as a Laplace transform \citep[12.13.112]{2007tisp.book.....G},
\begin{equation}\label{eq:id1}
    \frac{t}{s + \sqrt{s^2 - t^2}}
    = \int_{0}^{\infty} \! \frac{I_1(t k)}{k} \, \E^{-s k} \, \D k \;,
\end{equation}
which factorises the expressions in $s$ and $t$.
It remains to transform each factor into the corresponding integral on the left-hand side of the integral equation~\eqref{eq:h-eq}.
One is a simple Gaussian integral,
\begin{equation}\label{eq:id2}
    \E^{-s k} \, \E^{\frac{s^2}{4\sigma^2}}
    = \frac{1}{\sqrt{\pi} \, 2 \sigma} \int_{-\infty}^{\infty} \! \E^{-\frac{(z + 2 \sigma^2 k)^2}{4 \sigma^2}} \, \E^{\frac{z s}{2 \sigma^2}} \, \D z \;,
\end{equation}
the other is Weber's second exponential integral \citep[6.633.4]{2007tisp.book.....G},
\begin{equation}\label{eq:id3}
    I_1(t k) \, \E^{\frac{t^2}{2 \sigma^2}}
    = \frac{1}{\sigma^2} \, \E^{-\frac{\sigma^2 k^2}{2}} \int_{0}^{\infty} \! \E^{-\frac{r^2}{2 \sigma^2}} \, I_1(r k) \, I_1\Big(\frac{r t}{\sigma^2}\Big) \, r \, \D r \;.
\end{equation}
Inserting the identities~\eqref{eq:id1}--\eqref{eq:id3} into the right-hand side of the integral equation~\eqref{eq:h-eq} and changing the order of integration, we can simply read off the formal solution,\footnote{%
The same derivation holds, mutatis mutandis, for any variance~$\sigma_Z^2 \neq 2 \sigma^2$, which then yields a factor of $\sigma^2 + \sigma_Z^2$ instead of $3\sigma^2$ in the exponential.
}
\begin{equation}\label{eq:h-sn}
    h(r, z)
    = \int_{0}^{\infty} \! \E^{-\frac{3 \sigma^2 k^2}{2}} \, \frac{I_1(r k)}{r k} \, \E^{-z k} \, \D k \;.
\end{equation}
Since the integrand is well-defined everywhere, the solution exists and can, at least in principle, be calculated for every combination of values $r$ and $z$.
This is the unbiased ellipticity estimator from unweighted image moments.

Unfortunately, the integral~\eqref{eq:h-sn} appears to be missing from the usual tables, and we do not know whether an expression in closed form exists.
We therefore evaluate the function~$h$ numerically for given values of $r$ and $z$, which is detailed in Appendix~\ref{sec:computation}.

\subsection{Performance of the Unbiased Estimator}

Having found a way to accurately compute the unbiased ellipticity estimator, we can now analyse its performance as a function of true ellipticity~$\epsilon$ and signal-to-noise ratio~$\SNR$.
Fig.~\ref{fig:dist} shows the distribution of the estimates for the parallel ellipticity component introduced in Section~\ref{sec:estim}.
We find that with decreasing $\SNR$ and/or increasing~$|\epsilon|$, the distribution becomes heavily skewed to the right, with a long tail of ellipticity over-estimates along the true ellipticity direction.
This counteracts the movement of the bulk of probability mass towards the origin, with the distribution's mode and median increasingly far from the mean, which by construction remains at the correct (i.e.\@ unbiased) value.

\begin{figure}%
\centering%
\includegraphics[scale=0.66]{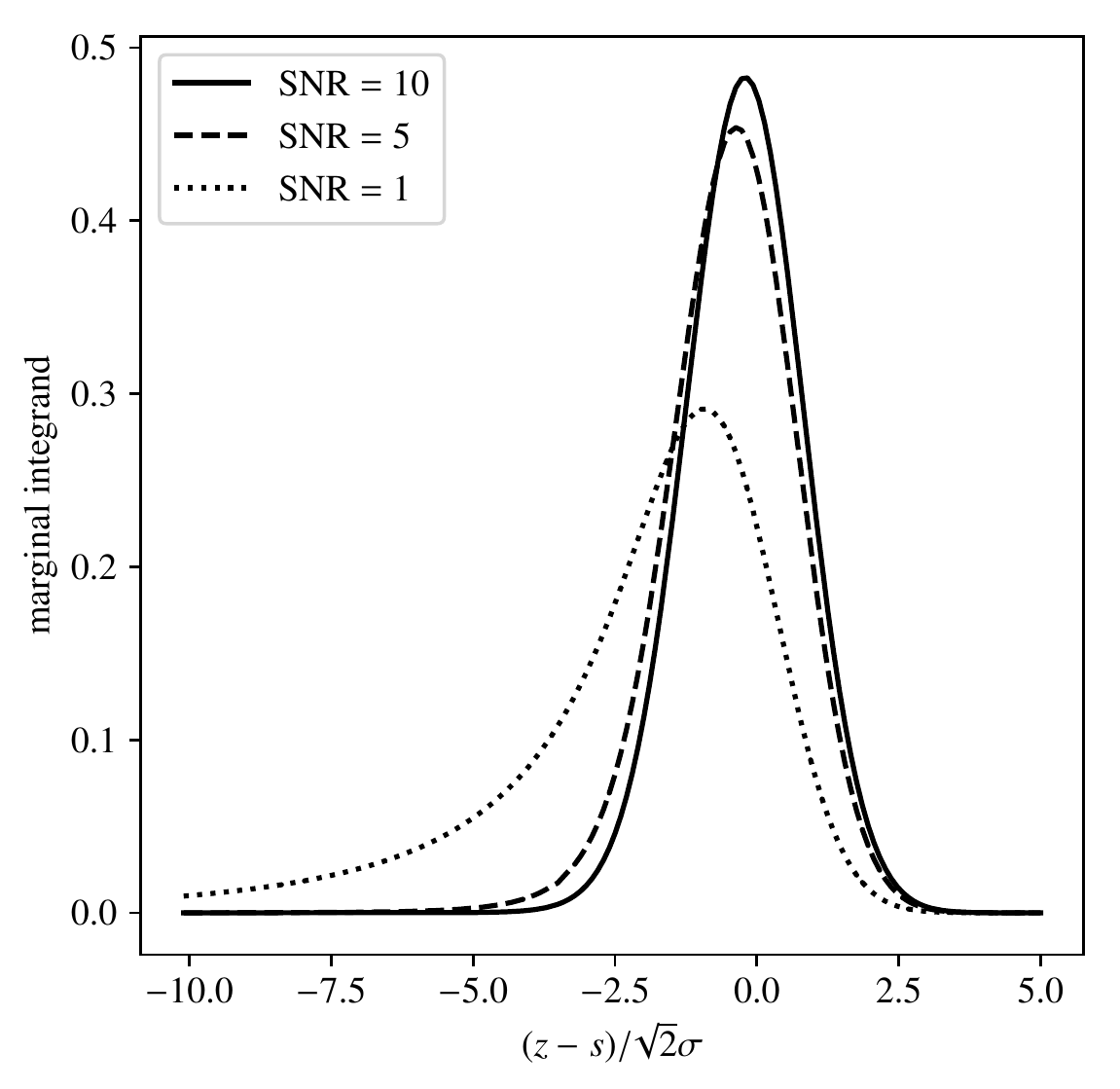}%
\caption{%
Marginal integrand of expectation~\eqref{eq:e-mar} for ellipticity~$|\epsilon| = 0.5$ at fixed signal-to-noise ratios~$\SNR$.
For high~$|\epsilon|$ or low SNR, the integrand decays very slowly, and events many standard deviations from the mean still contribute significantly to the expectation.
}%
\label{fig:marg}%
\end{figure}

The reason for the skewness and long tails of the distribution at low $\SNR$ or high $\epsilon$ is the dependency of the ellipticity estimator on increasingly rare events to become unbiased.
This is demonstrated by inserting the unbiased solution~\eqref{eq:h-sn} into the expectation~\eqref{eq:e-eps} of the generic estimator, carrying out the integration over~$r$ using identity~\eqref{eq:id3}, and writing the result in terms of the function~$h$, which we can calculate,
\begin{equation}\label{eq:e-mar}
    \Ev[\hat{\epsilon}]
    = \frac{\E^{\I \, \vartheta}}{\sqrt{\pi} \, 2 \sigma} \int_{-\infty}^{\infty} \! \sqrt{3/2} \, t \, h\big(\sqrt{3/2} \, t, \sqrt{3/2} \, z\big) \, \E^{-\frac{(z - s)^2}{4 \sigma^2}} \, \D z \;.
\end{equation}
When $\SNR \sim 1$ or $|\epsilon| \sim 1$ (i.e.\@ $t \sim s$), the integrand in marginal integral~\eqref{eq:e-mar} decays only slowly in the negative direction, where the rapid growth of the function~$h$ is no longer strongly suppressed by the Gaussian likelihood (Fig.~\ref{fig:marg}).
This means that increasingly unlikely events still contribute significantly to the expectation.

\begin{figure}%
\centering%
\includegraphics[scale=0.66]{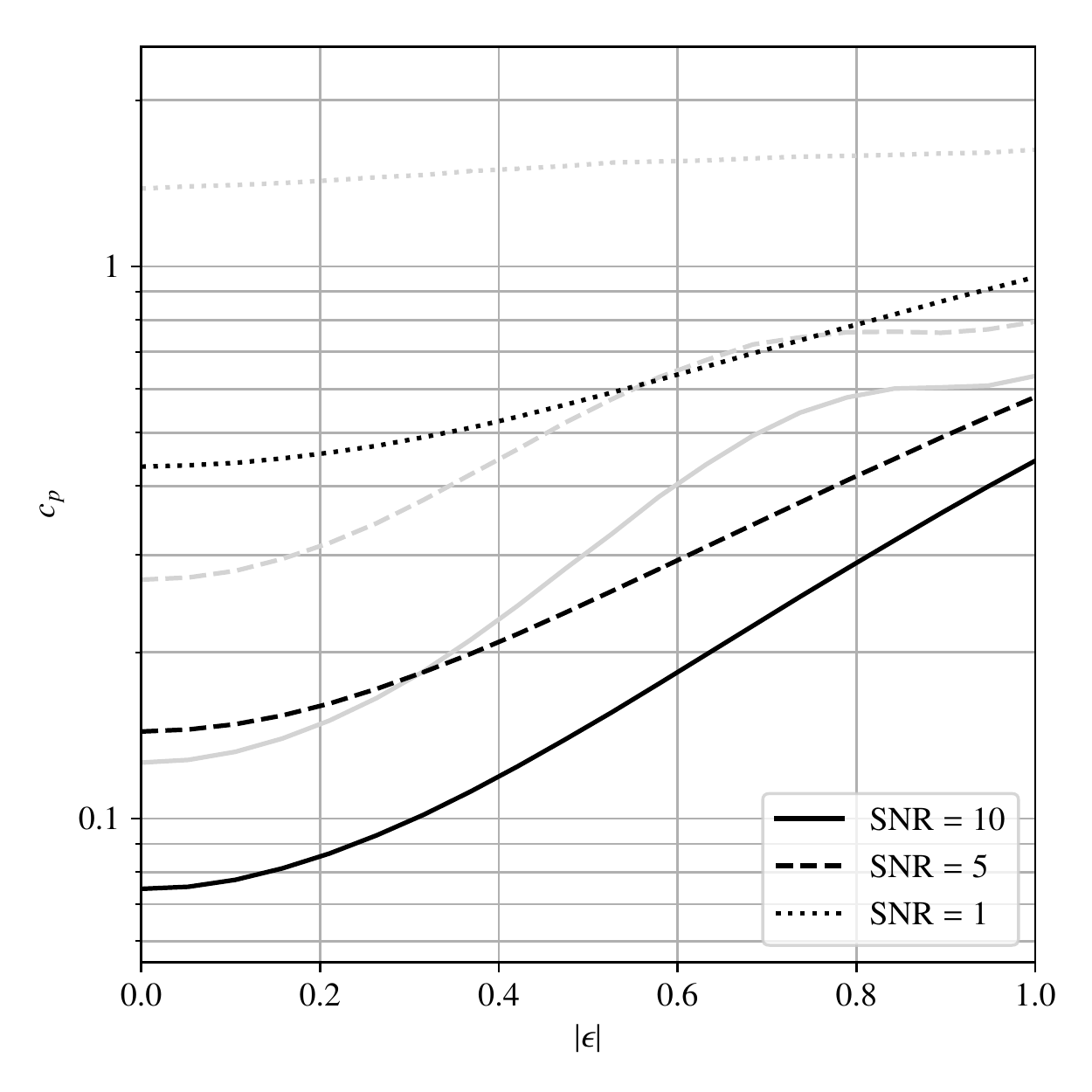}%
\caption{%
Confidence bounds~$c_p$ with $\Pr\big(|\hat{\epsilon} - \epsilon| \leq c_p\big) = p$ of the unbiased ellipticity estimator at the $p = 68\%$ (\emph{black}) and $p = 95\%$ (\emph{grey}) confidence levels as function of true ellipticity~$\epsilon$ for fixed signal-to-noise ratios~$\SNR$.
}%
\label{fig:conf}%
\end{figure}

It is then no surprise that the unbiased ellipticity estimator has infinite variance due to the divergence of the function~$h$ near infinity, similar to Voinov's estimator for the inverse mean \citep{Voinov:1985un}.\footnote{%
A further similarity is that the sample variance remains finite when $r$ and $z$ cannot become arbitrarily large, in contrast to e.g.\@ an estimator of the form~$1/z$, which diverges even for finite values as $z$ goes to zero.
}
However, the estimator has a well-defined confidence bound~$c_p$ at each confidence level~$p$,
\begin{equation}
    \Pr\big(|\hat{\epsilon} - \epsilon| \leq c_p\big) = p \;,
\end{equation}
and we use the confidence bounds at the 68 and 95 per cent level to tentatively quantify the accuracy of the estimator (Fig.~\ref{fig:conf}).
The values were obtained by randomly drawing $10^6$ samples of $X, Y, Z$ for given values of~$\epsilon$ and $\SNR$, and computing the $p$'th percentile of the absolute difference $|\hat{\epsilon} - \epsilon|$.

\begin{figure}%
\centering%
\includegraphics[scale=0.66]{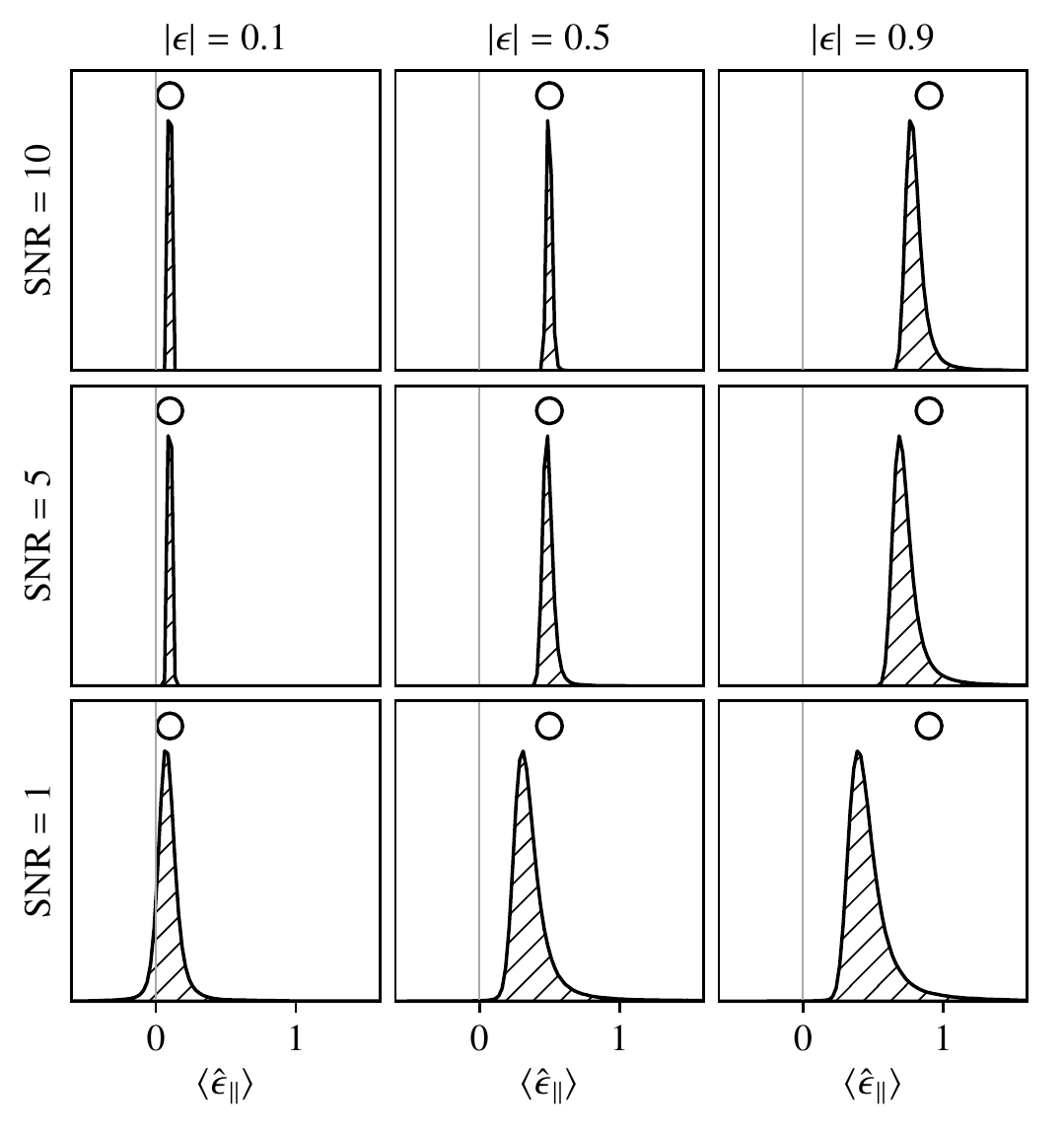}%
\caption{%
Distribution of the sample mean~$\langle \hat{\epsilon}_{\,\parallel} \rangle$ of 100 samples for the parallel component of the unbiased ellipticity estimator, together with the mean (\emph{circle}).
Due to the infinite variance of the estimator, the Central Limit Theorem does not hold, and the distributions remain skewed.
}%
\label{fig:mean}%
\end{figure}

Unfortunately, the infinite variance has practical consequences, as the Central Limit Theorem no longer holds, and the unbiased ellipticity estimator retains the skewness in the parallel component also in the distribution of the sample mean (Fig.~\ref{fig:mean}).
In applications where many individual ellipticity estimates are added together to obtain an estimate of  the shear, the result will not tend towards a normal distribution, which severely limits use of the unbiased ellipticity estimator for weak lensing.

\section{Discussion}
\label{sec:discussion}

The data reduction step from the observed image to the intrinsic Stokes parameters~\eqref{eq:u-sig}--\eqref{eq:s-sig} relies on five assumptions,
\begin{itemize}
\itemsep-3pt
	\item[(i)] that the intrinsic centroid of the source is known,
	\item[(ii)] that the moments of the PSF are known,
    \item[(iii)] that there is no contamination from neighbouring objects,
	\item[(iv)] that discretisation effects can be ignored, and
	\item[(v)] that the mask does not affect the signal.
\end{itemize}
To then take noise into account and derive the distribution of the observed random variables~\eqref{eq:X}--\eqref{eq:Z}, it was further assumed
\begin{itemize}
	\item[(vi)] that the pixel noise is (possibly correlated) Gaussian.
\end{itemize}
For real data, these assumptions are only approximately true, which can lead to additional biases beyond the scope of this work.
Some of these effects on moment-based ellipticity estimation have been investigated by \citet{2012MNRAS.424.2757M}.

If one accepts that the data reduction process indeed works, the results we obtain all follow from basic properties of sums of normal random variables.
In particular, that the Stokes parameters are the means of the random variables~\eqref{eq:X}--\eqref{eq:Z} follows directly from the fact that the noise has zero mean, and their covariance matrix~\eqref{eq:C} is that of any linear combination of normal random variables.
Multiple observations can be combined on the level of the random variables~$X, Y, Z$ simply because they probe the same parameters~$u, v, s$, modulo a possible rotation of the images.

Perhaps the only real surprise is that the covariance matrix~\eqref{eq:C-diag} of the relevant combinations of moments is close to diagonal, with a fixed ratio of $1 : 1 : 2$ for the individual variances.
However, this follows mostly from the symmetry of the moments itself, together with a suitable choice of mask for the pixels.
While it would certainly be necessary (and easily possible) to check this approximation in applications, we also describe a method to always achieve an exactly diagonal covariance matrix in Appendix~\ref{sec:covariance-fixing}.

After these preparatory considerations are in place, the actual process of ellipticity estimation with unweighted image moments is then relatively straightforward:
three independent random normal variables $X, Y, Z$ with known variances are observed; their means are the Stokes parameters~$u, v, s$; and the combination~\eqref{eq:eps} is the ellipticity~$\epsilon$, which is the parameter we ultimately wish to estimate.
We can deduce a number of general properties of this estimation problem, such as the signal-to-noise ratio~\eqref{eq:snr} and the Cram\'er-Rao lower bounds~\eqref{eq:crlb}--\eqref{eq:crlb-perp}.
Furthermore, we find that for the generic form~\eqref{eq:estim} of ellipticity estimators, the estimate is always unbiased in the ellipticity angle, a fact which has been noted in the literature \citep{2014MNRAS.439.1909V,1974ApJ...194..249W}.

To showcase the power of tackling the ellipticity estimation problem with a full statistical description, we were able to derive an unbiased ellipticity estimator in a few steps, and provide a thorough analysis of its performance.
This was done by simulating draws of the three random variables $X, Y, Z$ for given true ellipticity~$\epsilon$ and signal-to-noise level~$\SNR$, without the need for more complicated image simulations.
In the end, we found that this particular estimator is most likely unsuitable for use in actual cosmic shear surveys, due to the skewness of the distribution of the estimates, and its infinite variance.

Nevertheless, these results demonstrate the effectiveness of taking a statistical approach to deriving new results and estimators.
The fact that one only has to consider three independent normal random variables with known variance simplifies the process, to the point where one can work analytically, as demonstrated here.
Once a promising method has been developed, the restriction to using unweighted moments could even be dropped in applications.
The result is a biased ellipticity estimator, which can be calibrated as usual in cosmic shear surveys \citep[e.g.\@][]{2017MNRAS.465.1454H,2017arXiv170801533Z}.
Such a calibration would naturally also remove the additional biases mentioned at the beginning of this section.

\section{Conclusion}
\label{sec:conclusion}

Under very general assumptions, we have shown how the intrinsic Stokes parameters of a source can be recovered from observations that include effects such as a PSF and pixellation.
For Gaussian pixel noise, we then defined three jointly normal random variables with known covariance matrix, which is often close to diagonal (and can be made exactly diagonal by our proposed algorithm).
Since the Stokes parameters are the means of our set of random variables, measuring the ellipticity becomes a classical parameter estimation problem.
This provides a very useful framework with many readily available tools, such as the Cram\'er-Rao lower bound for the variance of an unbiased ellipticity estimator.

As an application of our method, we also derived the unbiased ellipticity estimator for shape measurement with image moments, which turned out to be impractical for weak lensing applications.
However, our results highlighted the importance of examining the full distribution of estimates, which can have undesirable properties even though some statistics (such as unbiasedness) would indicate perfect suitability for application to data.
Within our framework, this is easy and cheap to study because only uncorrelated Gaussian samples need to be drawn to test the estimator, instead of costly image simulations.

These results have therefore shown that unbiasedness may in practice not be the most desirable property of an ellipticity estimator.
Our statistical approach enables the straightforward construction and evaluation of new moment-based ellipticity estimators, which can be evaluated as demonstrated here (Wiegand et al.\@, in prep).
In the future, it might then be possible to pick and choose an ellipticity estimator that is precisely tailored to the problem at hand.

\section*{Acknowledgements}

We would like to thank Richard P. Rollins, Sean Holman, Martin Wiegand and Saralees Nadarajah for many helpful discussions.

The authors acknowledge support from the European Research Council in the form of a Consolidator Grant with number 681431.

\bibliographystyle{mnras}
\bibliography{medusa}

\appendix

\section{Moments of a Convolution}
\label{sec:moments-convolution}

To compute the required central moments of a convolution $f * g$ of functions $f$ and $g$, we begin by finding the raw moments,
\begin{multline}
    m_{pq}[f*g]
    = \int \! (f*g)(x, y) \, x^p y^q \, \D x \, \D y \\
    = \int \! f(x - x', y - y') \, g(x', y') \, x^p y^q \, \D x' \, \D y' \, \D x \, \D y \;.
\end{multline}
Writing $x^p y^q = (x - x' + x')^p (y - y' + y')^q$ and using the binomial theorem, the integrals can be factorised by a change of variables, and recover the individual raw moments of the functions~$f$ and $g$,
\begin{equation}\label{eq:m-fg}
    m_{pq}[f*g]
    = \sum_{k=0}^{p} \sum_{l=0}^{q} \binom{p}{k} \binom{q}{l} \, m_{kl}[f] \, m_{p-k,q-l}[g] \;.
\end{equation}
In particular, by direct application of the above to $m_{10}/m_{00}$, the centroid $(\bar x_{f*g}, \bar y_{f*g})$ of a convolution is
\begin{equation}\label{eq:xy-fg}
    (\bar x_{f*g}, \bar y_{f*g})
    = (\bar x_f + \bar x_g, \bar y_f + \bar y_g) \;,
\end{equation}
i.e.\@ the vectorial sum of the centroids $(\bar x_f, \bar y_f)$ and $(\bar x_g, \bar y_g)$ of the functions $f$ and $g$.

Furthermore, for any function~$\phi$, we can relate the central moments~$\mu_{pq}$ about the centroid to the raw moments by applying the binomial theorem to $(x - \bar x_\phi)^p \, (y - \bar y_\phi)^q$ in the integral,
\begin{multline}\label{eq:mu-m}
    \mu_{pq}[\phi]
    = \int \! \phi(x, y) \, (x - \bar x_\phi)^p \, (y - \bar y_\phi)^q \, \D x \, \D y \\
    = \sum_{k=0}^{p} \sum_{l=0}^{q} \binom{p}{k} \binom{q}{l} \, (-\bar x_\phi)^{p-k} \, (-\bar y_\phi)^{q-l} \, m_{kl}[\phi] \;.
\end{multline}
Using this relation for the convolution $f*g$, together with the raw moments~\eqref{eq:m-fg} and centroids~\eqref{eq:xy-fg} found above, and finally using the inverse of~\eqref{eq:mu-m} to reassemble the raw moments of~$f$ and~$g$ into central moments, we arrive at the second-order central moments of a convolution,
\begin{equation}
\begin{aligned}
    \mu_{20}[f*g] &= \mu_{20}[f] \, \mu_{00}[g] + \mu_{00}[f] \, \mu_{20}[g] \;, \\
    \mu_{11}[f*g] &= \mu_{11}[f] \, \mu_{00}[g] + \mu_{00}[f] \, \mu_{11}[g] \;, \\
    \mu_{02}[f*g] &= \mu_{02}[f] \, \mu_{00}[g] + \mu_{00}[f] \, \mu_{02}[g] \;,
\end{aligned}
\end{equation}
or, more concisely, $\mu_{pq}[f*g] = \mu_{pq}[f] \, \mu_{00}[g] + \mu_{00}[f] \, \mu_{pq}[g]$ for $p + q = 2$.
This is the relation used in Section~\ref{sec:signal}.

\section{Covariance Fixing}
\label{sec:covariance-fixing}

The covariance matrix~\eqref{eq:C} of the random variables~$X, Y, Z$ can be brought into any desired shape by a simple transformation:
Adding zero-mean Gaussian noise with covariance matrix~$\mat{\Theta}$ to the observed pixel data, the covariance matrix~$\mat{C}_+$ with added noise is
\begin{equation}
    \mat{C}_+
    = \mat{M} \, (\mat{\Sigma} + \mat{\Theta}) \, \mat{M}^\tp
    = \mat{C} + \mat{M} \, \mat{\Theta} \, \mat{M}^\tp \;,
\end{equation}
where $\mat{C}$ is the original covariance matrix of the unmodified random vector~$(X, Y, Z)$.
We can therefore fix the covariance matrix~$\mat{C}_+$ by solving the matrix equation
\begin{equation}\label{eq:cfix}
    \mat{M} \, \mat{\Theta} \, \mat{M}^\tp
    = \mat{C}_+ - \mat{C} \;.
\end{equation}
Since $\mat{\Theta}$ is the covariance matrix of the additional noise, it must be positive semi-definite (but not necessarily positive definite, since we might elect not to add any noise to some of the pixels).
Hence we can use the Cholesky decomposition to write $\mat{\Theta} = \mat{L} \mat{L}^\tp$ for some lower-triangular matrix~$\mat{L}$, and Eq.~\eqref{eq:cfix} becomes
\begin{equation}\label{eq:cfix-chol}
    \mat{M} \, \mat{L} \mat{L}^\tp \, \mat{M}^\tp
    = \mat{C}_+ - \mat{C} \;.
\end{equation}
The right-hand side is therefore $\mat{C}_+ - \mat{C} = \mat{A} \mat{A}^\tp$ for some matrix~$\mat{A}$, and the difference $\mat{C}_+ - \mat{C}$ must be a positive semi-definite matrix (i.e.\@ we cannot reduce the total variance by adding noise).

If the difference $\mat{C}_+ - \mat{C}$ is not positive (semi-)definite, the candidate matrix~$\mat{C}_+$ can be amended, e.g.\@ by a small increase in the diagonal elements, and tried again.
Practical tests for positive (semi-)definiteness are Sylvester's criterion \citep{1986JGCD....9..121P} or the success of a Cholesky decomposition of $\mat{C}_+ - \mat{C}$.
The latter method has the particular advantage that it produces the matrix~$\mat{A}$ directly.

Instead of solving Eq.~\eqref{eq:cfix} for~$\mat{\Theta}$, it then suffices to solve a smaller matrix equation for a reduced factor~$\mat{L}$,
\begin{equation}\label{eq:MLA}
    \mat{M} \, \mat{L} = \mat{A} \;,
\end{equation}
where~$\mat{L}$ now only has three columns (since~$\mat{A}$ is $3 \times 3$).
A solution is easily obtained using the pseudoinverse $\mat{M}^+ = \mat{M}^\tp(\mat{M} \mat{M}^\tp)^{-1}$,
\begin{equation}
    \mat{L} = \mat{M}^+ \mat{A} \;,
\end{equation}
since $\mat{M} \, \mat{L} = \mat{M} \, \mat{M}^\tp(\mat{M} \mat{M}^\tp)^{-1} \, \mat{A} = \mat{A}$ indeed solves Eq.~\eqref{eq:MLA}.
The pseudoinverse has the property that it produces the minimum-norm solution (for matrices: in the Frobenius norm), hence only the (in some sense) least amount of noise is added to the pixels.

With the matrix~$\mat{L}$ constructed, it suffices to draw a random vector~$\vec{N}$ of three standard normal variates and add the noise~$\mat{L} \, \vec{N}$ to the pixels:
The covariance matrix $\mat{\Theta} = \Cov[\mat{L} \, \vec{N}] = \mat{L} \mat{L}^\tp$ then indeed fulfils $\mat{M} \, \mat{\Theta} \, \mat{M}^\tp = \mat{M} \, \mat{L} \mat{L}^\tp \, \mat{M}^\tp = \mat{A} \mat{A}^\tp = \mat{C}_+ - \mat{C}$ as required by Eq.~\eqref{eq:cfix}, and the covariance matrix~$\mat{C}_+$ of the pixels with added noise is fixed to the desired form.

\section{Numerical Computation}
\label{sec:computation}

To compute the unbiased ellipticity estimator of Section~\ref{sec:ube}, it is necessary to numerically evaluate the integral form~\eqref{eq:h-sn} of the function~$h(r, z)$.
Inserting an integral representation for the modified Bessel function \citep[8.431.1]{2007tisp.book.....G},
\begin{equation}\label{eq:I-int}
    I_1(r k)
    = \frac{r k}{\pi} \int_{-1}^{1} \! \sqrt{1 - \xi^2} \, \E^{-r k \xi} \, \D \xi \;,
\end{equation}
the integration over~$k$ in the solution~\eqref{eq:h-sn} can be carried out.
The function~$h(r, z)$ is hence equivalently expressed as a finite integral,
\begin{equation}\label{eq:h-bi}
    h(r, z)
    = \frac{1}{\sqrt{6 \pi} \sigma} \int_{-1}^{1} \! \sqrt{1 - \xi^2} \, \E^{\frac{(r \xi + z)^2}{6 \sigma^2}} \erfc\Big(\frac{r \xi + z}{\sqrt{6} \sigma}\Big) \, \D \xi \;,
\end{equation}
where $\erfc(\,\cdot\,)$ is the complementary error function.\footnote{%
The product $\exp(x^2) \erfc(x)$ is sometimes available directly as the scaled complementary error function $\mathrm{erfcx}(x)$ to prevent under-/overflow.
}
In this form, the integral can be computed using Chebyshev-Gauss quadrature \citep[25.4.40]{1972hmfw.book.....A}, or a more general scheme for numerical integration.
However, depending on the combination of values for~$r$ and~$z$, a large number of function evaluations may be required before accurate results are obtained.

Instead, the modified Bessel function can be substituted by a series representation \citep[8.447.2]{2007tisp.book.....G},
\begin{equation}
    I_1(r k)
    = \sum_{n=0}^{\infty} \frac{1}{n! \, (n+1)!} \, \Big(\frac{r k}{2}\Big)^{2n+1} \;.
\end{equation}
The integration over~$k$ in the solution~\eqref{eq:h-sn} can then similarly be carried out \citep[3.462.1]{2007tisp.book.....G},
\begin{equation}
    h(r, z)
    = \frac{\E^{\frac{z^2}{12\sigma^2}}}{\sqrt{12} \sigma} \sum_{n=0}^{\infty} \frac{(2n)!}{n! \, (n+1)!} \, \Big(\frac{r}{\sqrt{12} \sigma}\Big)^{2n} \, D_{-2n-1}\Big(\frac{z}{\sqrt{3}\sigma}\Big) \;,
\end{equation}
where $D_\nu(\,\cdot\,)$ is the parabolic cylinder function.
This relates the function~$h$ to Voinov's estimator, or equivalently to the generalised Hermite functions $H_\nu(\,\cdot\,)$ of negative order,\footnote{%
For a normal random variable $X$ with mean $\mu$ and unit variance, the probabilists' Hermite polynomials $\mathit{He}_n(X)$ are the well-known unbiased estimators for non-negative integer powers $\mu^n$ of the mean.
The negative-order functions $\mathit{He}_{-n}(X)$ are precisely Voinov's estimator for $\mu^{-n}$.
}
\begin{equation}\label{eq:h-se}
    h(r, z)
    = \frac{1}{\sqrt{6} \sigma} \sum_{n=0}^{\infty} \frac{(2n)!}{n! \, (n+1)!} \, \Big(\frac{r}{\sqrt{6} \sigma}\Big)^{2n} \, H_{-2n-1}\Big(\frac{z}{\sqrt{6}\sigma}\Big) \;.
\end{equation}
The usual recurrence relation for Hermite polynomials continues to hold for negative orders and can be rearranged,
\begin{equation}\label{eq:H-rr}
    H_{-n-1}(x) = -\frac{x}{n} \, H_{-n}(x) + \frac{1}{2n} \, H_{-n+1}(x) \;,
\end{equation}
which requires the first negative-order Hermite function~$H_{-1}(x)$ to take the recursion past order zero,
\begin{equation}
    H_{-1}(x) = \frac{\sqrt{\pi}}{2} \, \E^{x^2} \erfc(x) \;, \quad
    H_0(x) = 1 \;.
\end{equation}
This follows directly from the results of \citet{Voinov:1985un}.
Repeatedly applying the relation~\eqref{eq:H-rr} to the series~\eqref{eq:h-se} then yields a recurrence relation for the $n$'th term~$S_n$,
\begin{equation}\label{eq:S-rr}
    S_n
    = \frac{z^2/\sigma^2 + 12n - 9}{36 \, n \, (n + 1)} \frac{r^2}{\sigma^2} \, S_{n-1}
    - \frac{2n - 3}{72 \, n^2 \, (n + 1)} \frac{r^4}{\sigma^4} \, S_{n-2} \;,
\end{equation}
with initial conditions given by the first negative-order Hermite function,
\begin{equation}
    S_0
    = \frac{1}{\sqrt{6}\sigma} \, H_{-1}\Big(\frac{z}{\sqrt{6} \sigma}\Big)
    = \frac{\sqrt{\pi}}{\sqrt{24} \sigma} \, \E^{\frac{z^2}{6 \sigma^2}} \erfc\Big(\frac{z}{\sqrt{6} \sigma}\Big) \;,
\end{equation}
and an additional term $S_{-1} = -z/r^2$ that does not appear in the series.
The function~$h$ can thus be evaluated efficiently to arbitrary precision by truncating the series
\begin{equation}
    h(r, z) = S_0 + S_1 + S_2 + \ldots
\end{equation}
once the remaining terms have fallen below a given threshold.

In practice, the forward recurrence~\eqref{eq:S-rr} is numerically unstable for positive values of~$z$.
In this case, Miller's algorithm can be used to compute the series via backward recursion \citep{Gil:2007km}.

\end{document}